# LETTER

# A cosmic web filament revealed in Lyman-$\alpha$ emission around a luminous high-redshift quasar

Sebastiano Cantalupo[1,2], Fabrizio Arrigoni-Battaia[3], J. Xavier Prochaska[1,2,3], Joseph F. Hennawi[3], & Piero Madau[1]

[1] Department of Astronomy and Astrophysics, University of California, Santa Cruz, CA, 1156 High Street, 95064, USA
[2] University of California Observatories, Lick Observatory, 1156 High Street, Santa Cruz, CA, 95064, USA
[3] Max-Planck-Institut für Astronomie, Königstuhl 17, D-69117 Heidelberg, Germany

**Simulations of structure formation in the Universe predict that galaxies are embedded in a "cosmic web"[1], where the majority of baryons reside as rarefied and highly ionized gas[2]. This material has been studied for decades in absorption against background sources[3], but the sparseness of these inherently one-dimensional probes preclude direct constraints on the three-dimensional morphology of the underlying web. Here we report observations of a cosmic web filament in Lyman $\alpha$ emission, discovered during a survey for cosmic gas fluorescently "illuminated" by bright quasars[4,5] at z~2.3. With a projected size of approximately 460 physical kpc, the Lyman-$\alpha$ emission surrounding the radio-quiet quasar UM287 extends well beyond the virial radius of any plausible associated dark matter halo. The estimated cold gas mass of the nebula from the observed emission - $M_{\rm gas} \sim 10^{12.0 \pm 0.5}/C^{1/2}$ solar masses, where C is the gas clumping factor – is at least ten times larger than what is typically found by cosmological simulations[5,6], suggesting that a population of intergalactic gas clumps with sub-kpc sizes may be missing within current numerical models.**

A recent pilot survey[5] using a custom-built, narrow-band (NB) filter on the Very Large Telescope (VLT) demonstrated that bright quasars can, like a flashlight, "illuminate" the densest knots in the surrounding cosmic web and boost fluorescent Lyman $\alpha$ emission[4,5,7–9] to detectable levels. Following the same experiment, we imaged UM287 on UT November 12 and 13, 2012 with a custom NB filter (NB3985) tuned to Lyman $\alpha$ at $z = 2.28$ inserted into the camera of the Low Resolution Imaging Spectrometer (LRIS) on the 10m Keck-I telescope (see Extended Data Figure 1). Figure 1 presents the processed and combined images, centered on UM287. In the NB3985 image, one identifies a very extended nebula originating near the quasar with a projected size of about 1 arcmin. Within the nebula, very few sources are identified in the broad-band images nor is any extended emission observed. This requires the NB light to be line-emission, and we identify it as Lyman-$\alpha$ at the redshift of UM287.

Figure 2 presents the NB3985 image, continuum subtracted using standard techniques (see Methods) and smoothed with a 1 arcsec Gaussian kernel. This image is dominated by the filamentary and asymetric Nebula that has a maximum projected extent of 55 arcsec measured from the $10^{-18}$ erg s$^{-1}$ cm$^{-2}$ arcsec$^{-2}$ isophotal, corresponding to about 460 physical kpc or 1.5 Mpc co-moving. Including (excluding) the emission from UM287 falling within the NB filter, the structure has a total line luminosity $L_{\rm Ly\alpha} = 1.43 \pm 0.05 \times 10^{45}$ erg s$^{-1}$ ($L_{\rm Ly\alpha} = 2.2 \pm 0.2 \times 10^{44}$ erg s$^{-1}$).

Although Lyman-$\alpha$ nebulae extending up to about 250 kpc have been previously detected[10–14], the UM287 Nebula represents so far a unique system: given its size, it extends well beyond any plausible dark matter halo associated with UM287 (see below) representing an exceptional example of emitting gas on truly Intergalactic scales.

The largest Lyman-$\alpha$ nebulae previously discovered (see Figure 3) are associated with the most massive dark matter haloes present in the high-redshift Universe. High-redshift Radio Galaxies (HzRGs), inferred to host obscured but luminous Active Galactic Nuclei (AGN)[11,13], are often surrounded by giant Lyman-$\alpha$ envelopes extending up to about 250 kpc at z~3[15]. Clustering arguments, the observation of large overdensities of Lyman $\alpha$ galaxies, together with the lack of X-ray detection from a possible intracluster medium, suggest that HzRGs are associated with $10^{13}$ M$_\odot$ haloes[15–17]. With a virial diameter of about 300 kpc at $z \sim 3$, these haloes are therefore able to contain the largest HzRG Lyman $\alpha$ nebulae. Blind NB surveys have derived an apparently different population of large nebulae (termed Lyman $\alpha$ blobs) with sizes extending up to 180 physical kpc at $z \sim 3$ that, in some cases, do not appear to be associated with a particular bright galaxy or AGN[12,14,18,19]. The rarity and the strong clustering of these sources, suggest, like for HzRG, an association with proto-cluster environments and haloes with masses of about $10^{13}$ M$_\odot$[20,21]. Although the detailed origin of the emission of the Lyman $\alpha$ blobs is still unclear, the sizes of the associated haloes strongly suggests that the emitting gas is confined within the halo itself. This is also the case for the Ly$\alpha$ nebulae previously detected around a small number of bright quasars, extending up to about 100 kpc[10,22–24]. Clustering studies demonstrate that bright quasars at $z < 3$ populate haloes of mass ~ $10^{12.5}$ M$_\odot$ - that have a virial diameter of about 280 kpc at $z \sim 2.3$ - independently of their redshift or luminosity[25,26].

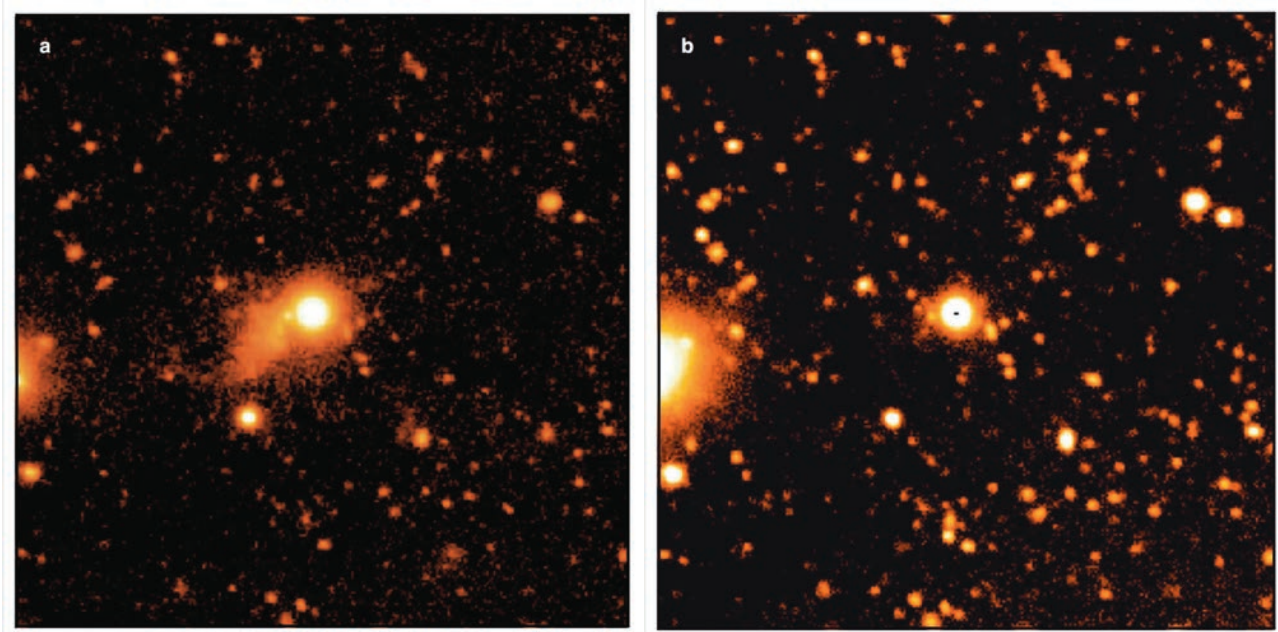

**Figure 1 | Processed and combined images of the field surrounding the quasar UM287.** Each image is 2 arcmin on a side and the quasar is located at the center. In the narrow-band (NB3985) image (panel "a"), which is tuned to the Lyman $\alpha$ line of the systemic redshift for UM287, one identifies very extended ($\approx$ 55 arcsec across) emission. The deep *V*-band image (panel "b") does not show any extended emission associated with UM287. This requires the Nebula to be line-emission, and we identify it as Lyman-$\alpha$ at the redshift of the quasar.

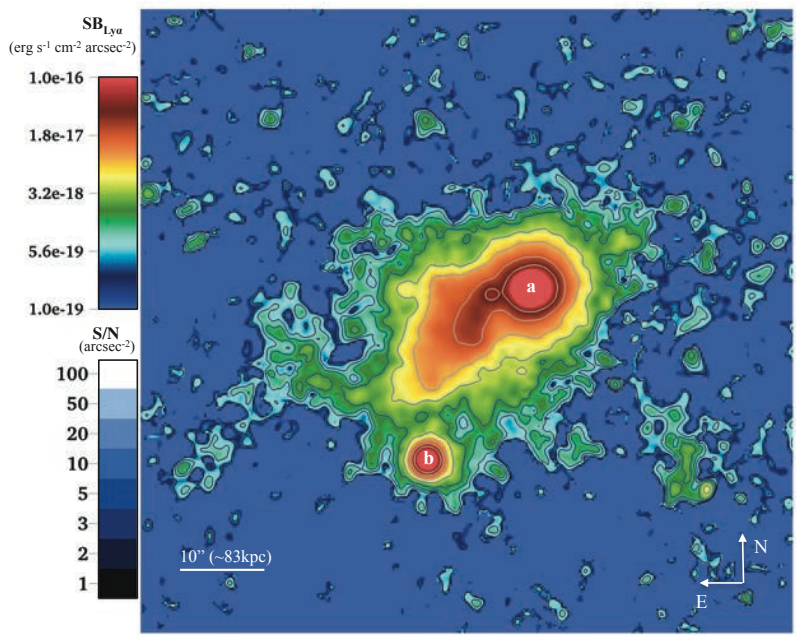

**Figure 2 | Lyman-$\alpha$ image of the UM287 Nebula.** We subtracted from the NB image the continuum contribution estimated from the broad-band images (see Methods). The location of UM287 is labeled with the letter "a". The color map and the contours indicates, respectively, the Lyman-$\alpha$ surface brightness and the signal-to-noise ratio (S/N) per arcsec$^2$ aperture. The extended emission spans a projected angular size of $\approx$ 55 arcsec (about 460 physical kpc), measured from the 2$\sigma$ ($\sim 10^{-18}$ erg s$^{-1}$ cm$^{-2}$ arcsec$^{-2}$) contours. Object "b" is an optically faint ($g\sim$23AB) quasar at the same redshift of UM287 (see Extended Data Figure 2). The Nebula appears broadly filamentary and asymmetric, extending mostly on the eastern side of quasar "a" up to a projected distance of about 35 arcsec (~285 physical kpc) measured from the 2$\sigma$ isophotal. The Nebula extends towards south-east in the direction of the faint quasar "b". However, the two quasars do not seem to be directly connected by this structure that continues as a fainter and spatially narrower filament. The large distance between the two quasars and the very broad morphology of the Nebula argue against the possibility that it may originate from an interaction between the quasar host galaxies (see Methods).

The exceptionality of the Nebula is not only due to its size - about 460 physical kpc – but also to the fact that it is associated with a radio-quiet quasar. These systems have the smallest host halo mass (~ $10^{12.5}$ M$_\odot$) and virial diameter (280 kpc) among previously detected objects and do not have radio-jets that may power Lyman-$\alpha$ emission on large scales[27]. In order to be fully contained within the virial radius of a dark matter halo centered on UM287, we would require a halo mass that is at least ten times larger than the typical value associated with radio-quiet quasars. This would make the host halo of UM287 one of the largest known at $z > 2$, a possibility that is excluded by the absence of a significant over density of Lyα emitters around UM287 compared to other radio-quiet quasars (see Methods). Differently from any previous detection, the Nebula is therefore an image of intergalactic gas at $z > 2$ extending beyond any individual, associated dark matter halo. The rarity of these systems may be explained by the combination of anisotropic emission from the quasars (typically only about 40% of the solid angle around a bright, high-redshift quasar is unobstructed[28]), the anisotropic distribution of dense filaments and light travel effects that, for quasar ages younger than a few Myr, further limit the possible "illuminated" volume.

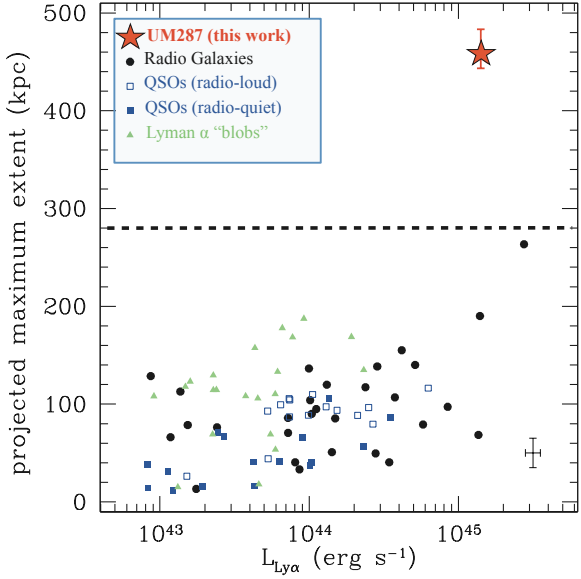

**Figure 3 | Luminosity-size relations for previously detected, bright Lyman-α nebulae and UM287.** The plot includes nebulae surrounding AGN and Lyman α blobs (LAB). The reported luminosities include the Lyman α emission (within the NB filters) from any sources embedded in the nebulae, if present. Excluding the contribution coming directly from the quasar Broad Line Region, the luminosity of the UM287 Nebula corresponds to $L_{Ly\alpha} = 2.2 \pm 0.2 \times 10^{44}$ erg s$^{-1}$ (about 16% of the total luminosity). Error bars for UM287 represent the 1σ photometric error including continuum-subtraction (errorbar is smaller than the symbol size) and an estimate on the error on the size measurement using ±1σ isophotal contours with respect to the $10^{-18}$ erg s$^{-1}$ cm$^{-2}$ arcsec$^{-2}$ isophotal. The typical errors for other sources are presented separately in the bottom-right corner. The dashed line indicates the virial diameter of a dark matter halo with total mass M ~ $10^{12.5}$ M$_\odot$, the typical host of radio-quiet quasars including UM287, as confirmed by the analysis of the galaxy overdensity in our field (see Methods). The UM287 Nebula, differently from any previous detection, extends on Intergalactic Medium scales that are well beyond any possible associated dark matter halo. Note that, even if we restrict the size measurement of the UM287 Nebula to the 4 × $10^{-18}$ erg s$^{-1}$ cm$^{-2}$ arcsec$^{-2}$ isophotal to be comparable with the majority of the previous surveys, the measured apparent size of the UM287 Nebula wil be reduced only by about 20%.

In order to constrain the physical properties of this, so far, unique system, we use a set of Lyman α radiative transfer calculations[29] combined with Adaptive Mesh Refinement (AMR) simulation of cosmological structure formation around a dark matter halo with mass $M_{DM}$ ~ $10^{12.5}$ M$_\odot$ (see Methods). We consider two possible, extreme scenarios for the Lyman α emission mechanism of the intergalactic gas associated with the Nebula: a) the gas is highly ionized by the quasar and the Lyman-α emission is mainly produced by hydrogen recombinations. b) the gas is mostly neutral and the emission is mainly due to scattering of the Lyman-α and continuum photons produced by the quasar Broad Line Region (BLR). The models are used to obtain the scaling relations between the observable Lyman-α surface brightness from the intergalactic gas surrounding the quasar and the hydrogen column densities (see Extended Data Figure 3). These scaling relations are consistent with analytical expectations. Note that the estimated column densities for case "a" are degenerate with the ionized gas clumping factor ($C = <n_e^2> / <n_e>^2$, where $n_e$ is the electron density) below the simulation resolution scale, ranging from ~ 10 proper kpc for diffuse intergalactic gas to ~ 160 pc for the densest regions within galaxies.

The results are presented in Fig.4. The observed Lyman-α emission requires very large column densities of "cold" ($T < 5 \times 10^4$ K) gas, up to $N_H$ ~ $10^{22}$ cm$^{-2}$. The implied total, cold gas mass "illuminated" by the quasar is $M_{gas} \approx 10^{12\pm0.5}$ M$_\odot$ for the "mostly ionized" case ("a") for $C = 1$ and $M_{gas} \approx 10^{11.4\pm0.6}$ M$_\odot$ for the "mostly neutral" case ("b"). Note that the total estimated mass for the case "a" scales as $C^{-1/2}$. For comparison, a typical simulated filament in our cosmological simulation of structure formation with size and morphology similar to the Nebula around a $M_{DM} \approx 10^{12.5}$ M$_\odot$ halo has a total gas mass of about $M_{gas} \approx 10^{11.3}$ M$_\odot$, but only about 15% of this gas is "cold" ($T < 5 \times 10^4$ K), i.e. $M_{gas} \approx 10^{10.5}$ M$_\odot$ and therefore able to emit substantial Lyman α emission. These estimates are consistent with a large sample of simulated haloes obtained by other recent works based on cosmological AMR simulations[6]. These simulations also show a (weak) decreasing trend of the cold gas fraction with halo mass.

How one can explain the large differences between the estimated, cold gas mass of the Nebula and the available amount of cold gas predicted by numerical simulations on similar scales? One possibility is to assume that the simulations are not resolving a large population of small, cold gas clumps within the low-density IGM that are illuminated and ionized by the intense radiation of the quasar. In this case, an extremely high clumping factor, up to $C$ ~ 1000, on scales below a few kpc would be required in order to explain the large luminosity of the Nebula with the cold gas mass predicted by the simulations. On the other hand, if some physical process that is not fully captured by current grid-based simulations increases the fraction of cold gas around the quasar, e.g. a proper treatment of metal mixing, a smaller clumping factor may be required. In the extreme - and rather unrealistic - case that all the hot gas is turned into a cold phase, the required clumping factor would be $C \sim 20$. Even if the gas is not ionized by the quasar (case "b" above), the simulations are able to reproduce the observed mass only if a substantial amount of hot gas is converted into a cold phase. Incidentally, this is exactly the same result found comparing the properties of Lyman α absorption systems around a large statistical sample of quasars with simulations[30]. Proper modeling of this gas phase will likely require a new generation of numerical models that are able - simultaneously - to spatially resolve these small IGM clumps within large simulation boxes, treat the multiphase nature of this gas and its interaction with galaxies and quasars.

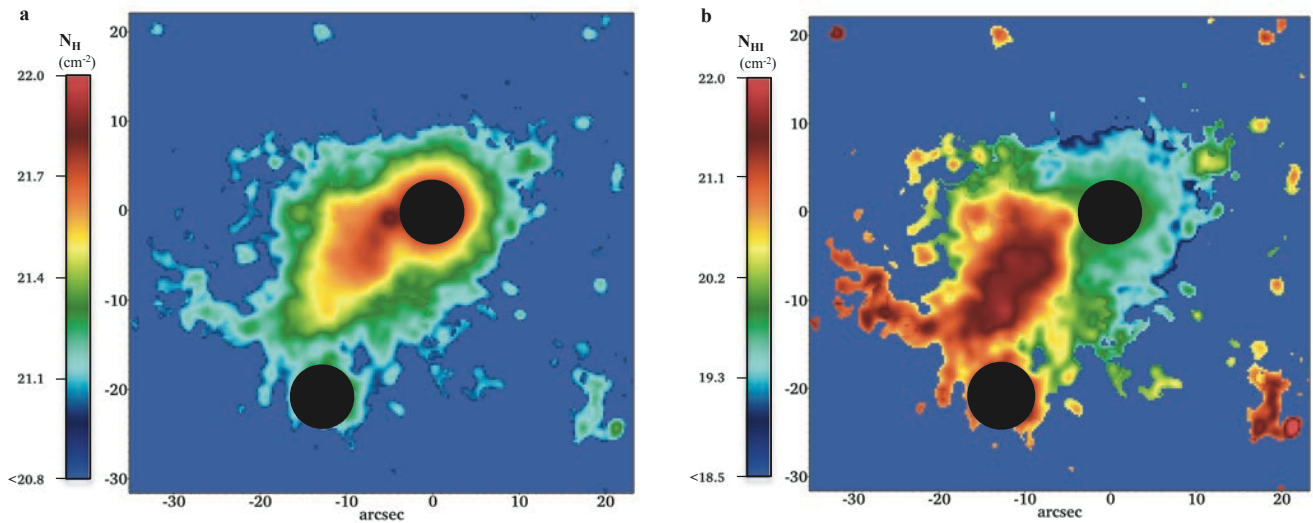

**Figure 4 | Inferred hydrogen column densities associated with the UM287 Nebula.** We have converted the observed Lyman-α SB into gas column densities using a set of scaling relations obtained with detailed radiative transfer simulations and consistent with analytical expectations (see Extended Data Figure 3 and Methods). We have explored two estreme cases: a) the gas is mostly ionized by the quasar radiation (panel "a"), b) the gas is mostly neutral (panel "b"). Two circular regions with a diameter of 7 arcsec (~ 8 times the seeing radius) have been masked at the location of the quasars (black circles). The inferred hydrogen column density in panel "a" scales as $C^{-1/2}$, where $C$ is the gas clumping factor below a spatial length of up to about 10 physical kpc at moderate overdensities (less than about 40 times the mean density of the Universe at $z = 2.28$). The implied column densities and gas masses, in both cases, are at least a factor of ten larger than what is typically observed within cosmological simulations around massive haloes, suggesting, e.g., that a large number of small clumps within the diffuse IGM may be missing within current numerical models.

**Methods Summary**

We observed UM287 for a total of 10 hours, in a series of dithered, 1200s exposures. In parallel, we obtained 10hrs of broad-band *V* images with the LRIS-red camera and 1hrs of *B*-band imaging. For all observations, we employed the D460 dichroic. We binned the blue CCDs 2×2 to minimize read noise. The images were processed using standard routines within the reduction software IRAF, including bias subtraction, flat fielding and illumination correction. A combination of twilight sky flats and unregistered science frames has been used to produce flat-field images and illumination corrections for each band. We have calibrated the photometry of our images using two spectrophotometric stars (Feige110 and Feige34) and the standard star field PG0231+051. To isolate the emission in the Lyman α line we estimated and then subtracted the continuum emission from discrete and extended sources contained within the NB3985 filter using a combination of the *V*-band and *B*-band. We derived a relation between the observable Lyman α emission from diffuse gas illuminated by a quasar and the gas column densities combining a Lyman α radiative transfer model with the results of a cosmological hydrodynamical simulation of structure formation at $z = 2.3$[5]. The cosmological simulation consists of a $40^3$ comoving Mpc$^3$ cosmological volume with a $10^3$ comoving Mpc$^3$ high resolution region containing a massive halo compatible with the expected quasar hosts ($M_{DM} \sim 10^{12.5}$ M$_\odot$). The equivalent base-grid resolution in the high-resolution region corresponds to a ($1024^3$) grid with a dark-matter particle mass of about $1.8 \times 10^6$ M$_\odot$. We used other additional 6 grid refinements levels, reaching a maximum spatial resolution of about 0.6 comoving kpc, i.e. about 165 proper pc at $z = 2.3$. We have then applied in post processing an ionization and Lyman-α radiative transfer using the RADAMESH Adaptive Mesh Refinement code[29].

**Acknowledgments**

We thank the staff of the W.M.Keck Observatory for their excellent support during the installation and testing of our custom-built narrow-band filter. S.C. thanks Martin Haehnelt for comments on an earlier version of the letter and Joel Primack for useful conversations. S.C. and J.X.P. acknowledge support from the National Science Foundation (NSF) grant AST-1010004. P.M. acknowledges support from the NSF through grant OIA-1124453, and from NASA through grant NNX12AF87G. The data presented herein were obtained at the W. M. Keck Observatory, which is operated as a scientific partnership among the California Institute of Technology, the University of California and the National Aeronautics and Space Administration. The Observatory was made possible by the generous financial support of the W. M. Keck Foundation. The authors wish to recognize and acknowledge the very significant cultural role and reverence that the summit of Mauna Kea has always had within the indigenous Hawaiian community. We are most fortunate to have the opportunity to conduct observations from this mountain.

**Author contribution statements**

S.C. designed the observational survey and the custom-built filter, conducted the observations, led the narrow-band imaging data reduction and analysis, performed the numerical simulations and led the theoretical interpretation, the writing of the text and the production of the figures. F.A.B. and J.X.P. assisted with the observations, contributed to data reduction, the text and the figures. In particular, F.A.B. reduced and calibrated the images, produced the continuum-subtracted image, the catalogues of LAEs, and compiled data on all Lyman-α nebulae in the published literature. J.X.P. reduced the spectrum of the companion quasar and contributed to the text. J.F.H. and P.M. contributed to the text and assisted with the planning and interpretation of the observations.

**Corresponding Author**

Correspondence and requests for materials should be addressed to S. Cantalupo, cantal@ucolick.org.

# METHODS

## Observations and data reduction

As part of an ongoing program to search for Lyman-$\alpha$ emission associated with the fluorescence of quasar ionizing radiation[5], we obtained deep, narrow-band (NB) imaging of the field surrounding UM287, also known as PHL 868 and LBQS 0049+0045. UM287 was discovered in the University of Michigan emission-line survey, has a precisely measured redshift $z = 2.279 \pm 0.001$ based on analysis of [OIII] emission lines[31], and has a Bolometric luminosity $L_{Bol} \approx 10^{47.3}$ erg s$^{-1}$ estimated from its 1450Å rest-frame flux using standard cosmology[32]. This places it in the upper quartile of UV-bright quasars at this redshift. Assuming that the spectral energy distribution follows a power-law[33] ($f_\nu = \nu^{-1.57}$) at energies exceeding 1 Ryd, we estimate the luminosity of ionizing photons[34] to be $\Phi = 10^{57.3}$ s$^{-1}$ assuming isotropic emission.

The quasar has no counterpart in the FIRST[35] images at 20cm (1.4 GHz), and based on the FIRST coverage maps we obtain a 5$\sigma$ limit of $F_{radio} < 0.76$ mJy, which, given its large UV luminosity, classifies this quasar as radio-quiet[36]. We selected this source for imaging based solely on its high luminosity, its precisely measured redshift, and its radio-quiet characteristics. We purchased a custom-designed NB filter from Andover Corporation, sized to fit within the grism holder of the Keck/LRISb camera. The filter was tuned to Lyman $\alpha$ at the source's systemic redshift and we requested a narrow band-pass (FWHM ~ 3nm) that minimized sky background while maximizing throughput. Extended Data Figure 1 presents the as-measured transmission curve of the NB3985 filter.

We observed UM287 on the nights of UT 12-13 November 2013 for a total of 10 hours, in a series of dithered, 1200s exposures. Conditions were clear with atmospheric seeing varying from FWHM $\approx$ 0.6 to 1 arcsec. In parallel, we obtained 10hrs of broad-band $V$ images with the LRISr camera and 1hrs of $B$-band imaging. For all observations, we employed the D460 dichroic. We binned the blue CCDs 2×2 to minimize read noise.

All of these data were processed with standard techniques. Bias subtraction was performed using measurements from the overscan regions of each image. The images have been reduced using standard routines within the reduction software IRAF, including bias subtraction, flat fielding and illumination correction. A combination of twilight sky flats and unregistered science frames has been used to produce flat-field images and illumination corrections for each band. Each individual frame has been registered on the SDSS-DR7 catalogue using in SExtractor[37] and SCAMP[38] in sequence. The astrometric uncertainty of our registered images is about 0.2 arcsec. Finally, for each band (NB3985, $B$, $V$), the corrected frames were average-combined using SWarp[39].

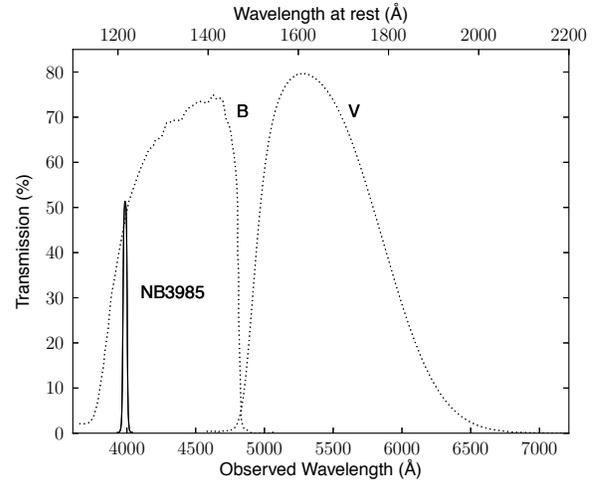

**Extended Data Figure 1 | Measured transmission curves of the filters used in this study.** The top axis indicates the rest-frame wavelength for sources at z = 2.27.

We have calibrated the photometry of our images in the following manner. First, we observed during the two nights two spectrophotometric stars (Feige110 and Feige34) through the NB filter, under clear conditions. For the broad band images, we observed the standard star field PG0231+051.

To compute the zero-point for the narrow-band images, we first measured the number of counts per second of the standard stars Feige110 and Feige34. We then compared this measurement with the flux expected, estimated by convolving the spectrum of the standard star with the normalized filter transmission curve (Extended Data Figure 1). The two measurements agreed to within 0.1 mag. We attribute the difference to small variations in the transparency and adopt an average zero-point of 24.14 mag. The surface brightness limit for our observation in the central region of the image occupied by the Nebula is about $5 \times 10^{-19}$ erg s$^{-1}$ cm$^{-2}$ arcsec$^{-2}$ at 1$\sigma$ level within an aperture of 1 arcsec$^2$.

For the broad-band images, we compared the number of counts per second of the five stars in the PG0231+051-field with their tabulated $V$ and $B$ magnitudes[40]. The derived zero-point for the five stars are consistent with each other within a few percent and we adopt the average values: $B_{ZP} = 28.40$ and $V_{ZP} = 28.07$.

As the standard stars and the PG0231+051-field were observed with a similar airmass of approximately 1.2, which corresponds to the average airmass of our observations, we did not correct the individual images before combination. Moreover, by monitoring unsaturated stars on several exposures, we estimated that the correction would be of the order of a few percent.

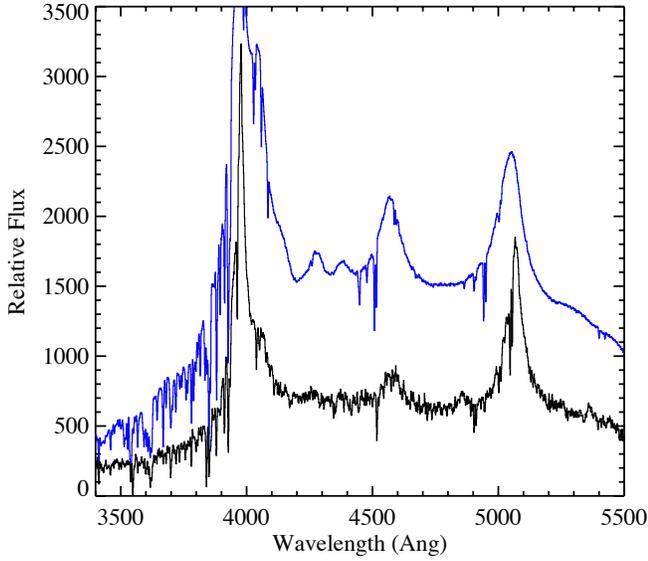

**Extended Data Figure 2 | Keck/LRIS spectrum of the faint, radio-loud companion quasar of UM287.** This object is indicated by the letter b in Figure 2 and separated by about 24 arcsec from UM287 (black line). This spectrum clearly shows that this source is a quasar at a similar redshift with respect to UM287 (blue line).

**Continuum Subtraction**

To isolate the emission in the Lyman-α line we estimated and then subtracted the continuum emission from discrete and extended sources contained within the NB3985 filter. We estimated the continuum using a combination of the *V*-band and *B*-band images as follows. Firstly, we smoothed both of the broad-band images using a Gaussian kernel of 1 arcsec and set to zero all of the pixels with values less than the measured root-mean-square (1σ). Additionally, in the *V*-band we set to zero all of the pixels which have signal above 1σ in the *B*-band as we prefer to use the latter image when possible given that it lies closer in wavelength to the Lyman-α line.

After matching the seeing between the narrow-band and the broad-band images, the continuum subtraction has been applied using the following formula

$$Ly\alpha = \text{NB3985} - a \cdot \left(\frac{\text{FWHM}_{\text{NB3985}}}{\text{FWHM}_B}\right)\left(\frac{Tr_{\text{NB3985}}}{Tr_B}\right)\cdot B - b \cdot \left(\frac{\text{FWHM}_{\text{NB3985}}}{\text{FWHM}_V}\right)\left(\frac{Tr_{\text{NB3985}}}{Tr_V}\right)\cdot V$$

where *Lyα* is the final subtracted image, NB3985 is the smoothed narrow-band image, *B* and *V* are the smoothed and masked broad-band images, and $Tr_{\text{NB3985}}$, $Tr_B$ and $Tr_V$ are the transmission peak values for NB3985, *B*-band and *V*-band filters, respectively. The parameters $a = 0.85$ and $b = 0.65$ allow a better match to the continuum. Following this procedure, we primarily used the smoothed *B*-band image to estimate the continuum and we included the *V*-band to achieve deeper sensitivity and to correct those objects not detected in the *B*-band image.

**Data reduction and analysis for the companion quasar**

Upon analyzing the continuum-subtracted Lyman α image, we identified a compact Lyman-α excess source at 24.3 arcsec separation from UM287 (corresponding to about 200 physical kpc), which has a faint counterpart in our LRIS continuum image and is also detected in the SDSS ($g = 22.8 \pm 0.1$). Further exploration of this source reveals it is detected by the FIRST survey (FIRSTJ005203.26+010108.6) with a flux $F_{\text{peak}} = 21.38$ mJy, strongly suggesting that this source is a radio-loud but optically faint quasar. On UT 08 December 2013, we obtained a long-slit spectrum of J005203.26+010108.6 using the Keck/LRIS spectrometer configured with the D560 dichroic, the 600/4000 grism in the LRISb camera, and the 600/10000 grating in the LRISr camera. We oriented the longslit to also cover UM287. These data were reduced with the LowRedux (http://www.ucolick.org/~xavier/LowRedux/index.html) software package using standard techniques. Extended Data Figure 2 presents the two, optimally extracted spectra from the LRISb camera. One recognizes the broad and bright emission lines characteristic of Type I quasars. The redshift estimated from these lines - that has an error of about 800 km s$^{-1}$ (1σ) - is consistent with the systemic redshift of UM287, suggesting that UM287 is actually a member of a binary system with a fainter companion. We emphasize, however, that there is very little (if any) Lyman α emission apparent in the NB image that may be associated with J005203.26+010108.6 aside from that produced by its own nuclear activity.

Because of the large distance from UM287 - at least 200 proper kpc and up to 4 proper Mpc considering the 1σ redshift error, and the morphology of the Nebula we can exclude that the UM287 Nebula is the result of tidal interaction due to a merging event between the two quasar hosts. Indeed, such a large separation would imply that any possible encounter between the two quasars is likely a high velocity interaction or an encounter with large impact parameter. We note that it is not impossible but extremely difficult to produce a long and massive tidal tail during a "fast" encounter[41] and the amount of gas stripped by the quasar host galaxies in the best scenario would likely be a very small fraction (< 10%) of its total ISM. Irrespectively of the details of the possible interaction between the two quasar host galaxies, any resulting, long tidal tail would be very thin with sizes of the order of few kpc or less[41] while the observed Nebula has a FWHM thickness of at least 100 physical kpc in its widest point.

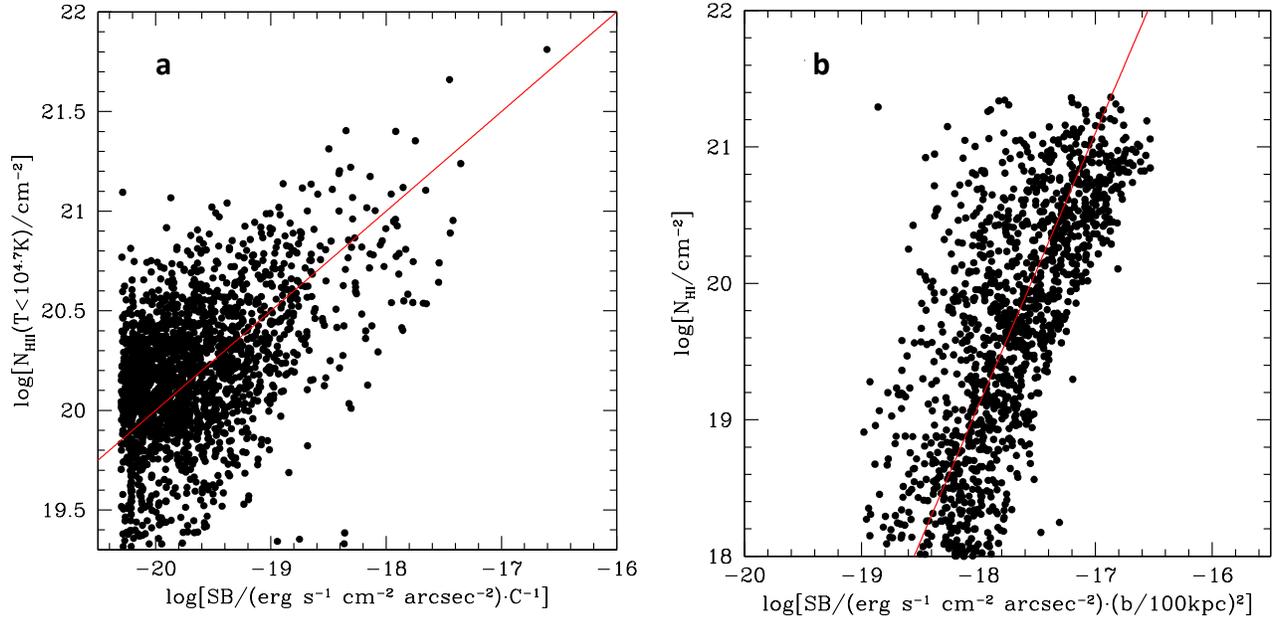

**Extended Data Figure 3 Pixel to pixel correlations for Lyman α SB for case "a" and case "b".** Panel "a": pixel to pixel correlation between simulated Lyman α SB divided by the clumping factor (C) and corresponding cold ($T < 5 \times 10^4$ K) ionized hydrogen column densities for case "a" (see text for details). The solid line indicates the relation: $N(\text{HII}) = 10^{21} \times [\text{SB} / (10^{-18} \text{ erg s}^{-1} \text{ cm}^{-2} \text{arcsec}^{-2})]^{1/2} \times C^{-1/2}$ cm$^{-2}$. Panel "b": pixel to pixel correlation between simulated Lyman α SB (normalized by the quasar impact parameter squared, $b^2$) and corresponding neutral hydrogen column density for case "b" (see text for details). The solid line represents the relation: $N(\text{HI}) = 10^{19.1} \times [\text{SB} / (10^{-18} \text{ erg s}^{-1} \text{ cm}^{-2} \text{ arcsec}^{-2}) \times (b / 100 \text{ kpc})^2]^2$ cm$^{-2}$.

### Galaxy overdensity analysis

We have obtained a sample of 60 Lyman α emitter (LAE) candidates above a flux limit of $3 \times 10^{-18}$ erg s$^{-1}$ cm$^{-2}$ (corresponding to a Lyman α luminosity of about $2 \times 10^{41}$ erg s$^{-1}$) within the volume probed by our NB imaging (~ 3100 comoving Mpc$^{-3}$) around UM287. The selection is based on the same technique applied to our pilot survey[5].
How does the number density in our survey compare to other similar searches around massive objects? Surveys of LAE around HzRG[42,15] have revealed large overdensities of LAEs with respect to field studies at similar redshifts[43,44] that are compatible with the presence of a massive halo as estimated from clustering, e.g. $10^{13}$ M$_\odot$. NB imaging of the radio-galaxy MCR 1138-262 at $z = 2.16$ [42], associated with a 200 kpc scale Lyman α nebula, found a number density of LAE above $L_{\text{Ly}\alpha} = 1.4 \times 10^{42}$ erg s$^{-1}$ of $n_{\text{HzRG}} \sim 10 \pm 2 \times 10^{-3}$ comoving Mpc$^{-3}$ [15]. By comparison, the number density of LAE above the same limit at the same redshift in the field is $n_{\text{field}} \sim 1.5 \pm 0.5 \times 10^{-3}$ comoving Mpc$^{-3}$, corrected for completeness[43]. If we restrict our sample to the same luminosity cut we found a number density of $n_{\text{UM287}} \sim 5 \pm 1 \times 10^{-3}$ comoving Mpc$^{-3}$. Note that, at this luminosity, we are essentially complete. Despite the large statistical errors, we note that the overdensity with respect to the field around UM287 (about a factor three) is significantly smaller than the overdensity of LAE around MCR 1138-262 (about a factor six). A similar result is obtained comparing the overdensity of LAE around UM287 with other HzRG[15] suggesting that UM287 is hosted by a smaller halo than typical HzRG hosts.

Moreover, the modest overdensity of our field is strong evidence against the possibility that the UM287 Nebula may be fully contained by an individual dark matter halo of mass $10^{13.5}$ M$_\odot$, as would be required by its size. Note that the galaxy number density estimate around UM287 is a conservative upper limit: if the quasar is illuminating the surrounding volume, we expect a boost in the number of detectable Lyman α emitting objects due to fluorescence, as demonstrated in our pilot survey[5]. Our measurement is also compatible with the number density of LAEs found by other recent, shallower surveys for Lyman α emission around eight radio-quiet, bright quasars[45] at $z \sim 2.7$ that have a host halo mass of $10^{12.5}$ M$_\odot$ as constrained by the clustering of Lyman Break Galaxies.
These studies found number densities ranging from $6 \times 10^{-3}$ to $22 \times 10^{-3}$ comoving Mpc$^{-3}$ around individual quasars above a Lyman α luminosity of $L_{\text{Ly}\alpha} = 5.8 \times 10^{41}$ erg s$^{-1}$. Combining the 8 fields, the average number density from their survey is $12.0 \pm 0.4 \times 10^{-3}$ comoving Mpc$^{-3}$. Using the same luminosity cut, we find a number density of $12 \pm 2 \times 10^{-3}$ comoving Mpc$^{-3}$, suggesting that the halo mass of UM287 is indeed within the typical range for the host halos of radio-quiet quasars.

### Converting the observed Lyman-α emission to gas column densities

We derived a relation between the observable Lyman-α emission from diffuse gas illuminated by a quasar and the gas column densities combining a Lyman-α radiative transfer model with the results of a cosmological hydrodynamical simulation of structure formation at $z = 2.3$ [5]. The cosmological simulations have been obtained with the Adaptive Mesh Refinement code RAMSES[46] and consist of a $40^3$ comoving Mpc$^3$ cosmological volume with a $10^3$ comoving Mpc$^3$ high resolution region containing a massive halo compatible with the expected quasar hosts ($M_{DM} \sim 10^{12.5}$ $M_\odot$). The equivalent base-grid resolution in the high-resolution region corresponds to a ($1024^3$) grid with a dark-matter particle mass of about $1.8 \times 10^6$ $M_\odot$. We used other additional 6 grid refinements levels, reaching a maximum spatial resolution of about 0.6 comoving kpc, i.e. about 165 proper pc at $z = 2.3$. Star formation, supernova feedback, and an optically thin UV background with an on-the-fly self shielding correction are included using a typical choice of sub-grid parameters for the simulation resolution[5]. We have then applied in post processing an ionization and Lyman-$\alpha$ radiative transfer using the RADAMESH Adaptive Mesh Refinement code[29]. Ionization, Lyman-$\alpha$ and non-ionizing continuum radiation from the quasar Broad Line Region (BLR) is propagated within two symmetric cones that cover half of the solid angle around the quasar. We included lighttravel and finite light-speed effects for both ionizing and Lyman $\alpha$ radiation transfer and varied the quasar age (from 1 Myr to 10 Myr) and the orientation of the emission cones with respect to the observer line-of-sight and the cosmic web surrounding the simulated halo. We note that these effects are able to produce asymmetric Lyman-$\alpha$ nebulae with sizes and morphologies similar to the observations for short quasar ages (< 5 Myr).

In order to produce a calibrated relation for the case "a" as discussed in the main text, we have fixed the quasar ionizing and Lyman $\alpha$ luminosity to the observed value and assumed that the ionizing and Lyman $\alpha$ emitting cones are coincident. We have then produced mock images with the same angular resolution of the observation that have been convolved with a Point Spread Function (PSF) with 1 arcsec size to simulate atmospheric seeing. A column density map of cold ($T < 5 \times 10^4$ K) ionized hydrogen was produced from the simulations considering only the gas "illuminated" by the quasar and convolved with the same PSF. We have then cross-correlated the two quantities pixel by pixel and fitted the calibrated relation shown as a solid line in the left panel of Extended Data Figure 3. This relation is consistent with analytical expectations from highly ionized gas where the Lyman-$\alpha$ emission is mostly produced by hydrogen recombinations with a negligible contribution from collisional excitations and Lyman-$\alpha$ scattering (or photon-pumping) from the quasar non-ionizing continuum and Lyman-$\alpha$ radiation[5]. We have repeated the experiment varying the sub-grid clumping factor (C) below the simulation resolution and obtained, as expected for highly ionized gas, that the simulated SB scales linearly with C at a given gas column densities.

We have also considered the extreme case in which the simulated gas is only invested by non-ionizing radiation from the quasar, and therefore that dense gas in the simulation remains mostly neutral (case "b" in the main text) above the self-shielding density to the cosmic UV background (about 0.01 atoms cm$^{-3}$). We obtained and post-processed a mock image as in the previous case and cross-correlated the resulting Lyman-$\alpha$ SB with the neutral hydrogen column densities ($N_{HI}$). Despite the large scatter, we found a good correlation between these two quantities (right panel of Extended Data Figure 3) if the SB is normalized by the impact parameter (b) squared. The relation between the Lyman-$\alpha$ SB, neutral hydrogen column density and impact parameter is consistent with simple analytical expectations from pure scattering from the BLR of the quasar for Lyman-$\alpha$ optical depth much larger than unity. In this case, the amount of photon-pumping (or, analogously, the equivalent width of the absorbed quasar Lyman-$\alpha$ and continuum emission) is dominated by the line damping wing and therefore it is proportional to $N_{HI}^{1/2}$.